\documentclass[twocolumn]{revtex4}
\usepackage{graphicx}
\usepackage{dcolumn}
\usepackage{amsmath}

\begin{document}
\title{Logarithmic diffusion and porous media
equations: a unified description}
\author{ \footnote{E-mail adress: thamar@unioeste.br} I. T. Pedron$^1$, R. S. Mendes$^2$, \footnote{\emph{In
memoriam}}T.~ J. ~Buratta$^2$, L. C. Malacarne$^2$, and E. K.
Lenzi$^2$}
 \affiliation{ $^1$ Universidade
Estadual do Oeste do Paran\'a, Rua
Pernambuco, 1777, 85960-000, Marechal C\^andido Rondon, Paran\'a, Brazil\\
$^{2}$ Departamento de F\'\i sica, Universidade Estadual de
Maring\'a, Avenida Colombo, 5790, 87020-900, Maring\'a, Paran\'a,
Brazil }


  %\date{\today}

% ----------------------------------------------------------------

\begin{abstract}
{In this work we present the logarithmic diffusion equation as a
limit case when the index that characterizes a nonlinear
Fokker-Planck equation, in its diffusive term, goes to zero. A
linear drift  and a source term are considered in this equation. Its
solution has a lorentzian form, consequently this equation
characterizes a super diffusion like a L\'evy kind. In addition is
obtained an equation that unifies the porous media and the
logarithmic diffusion equations, including a generalized diffusion
equation in fractal dimension. This unification is performed in the
nonextensive thermostatistics context and increases the
possibilities about the description of anomalous diffusive
processes.}
\end{abstract}

\maketitle

\section{Introduction}

Diffusion is a very usual phenomena in the nature and it happens,
in general, when the system goes to an equilibrium state .
Therefore it is of fundamental relevancy in physics, chemical, and
biological processes.
 The linear dependence in the time growth of the  mean square
 displacement   $\langle x^2(t)\rangle \propto t$, or alternatively in the
 variance  (when $\langle x \rangle \neq 0 $), is the finger point
 of the Brownian movement and usual
 diffusion. It is a direct consequence of the central limit
 theorem and the  Markovian nature of the underlying stochastic process
  \cite{ma}. In contrast, the anomalous diffusion is in general
characterized by a non linear variance growth in the time,  that is
to say, the diffusion will be considered anomalous when the
behavior goes out the former.  This kind of diffusion  has a
fundamental role in the analysis of a large class of systems such as
plasma diffusion \cite{berryman}, diffusion in turbulent fluids
\cite{klafter86,klafter87}, fluids transportation in porous media
 \cite{spohn}, chaotic dynamics
\cite{klafter93}, non-gaussian behavior of the heartbeat
\cite{cpeng},  diffusion on fractals \cite{procaccia1,
procaccia2,stephenson,walter}, anomalous diffusion at liquid
surfaces
 \cite{bychuk},  in the study of vibrational energy in
proteins \cite{Yu}, among other physical systems. In the anomalous
behavior description the variance growth can be a power law kind
$\langle (\Delta x)^{2}\rangle \propto t^{\eta}$, or even present
another pattern. In this classification, when $\eta >1$, we have a
superdiffusive   process,  $\eta < 1$,  a subdiffusive one and $\eta
=1$ describes  usual diffusion. Furthermore, in an anomalous
diffusive process, the variance can be not finite, describing a
L\'evy process, although it presents a well defined index that
characterizes such process \cite{georges}. The description of this
kind of processes is based in the validity of the generalized
central limit theorem, termed L\'evy-Gnedenko, which states that, by
$N$-fold convolution, a distribution with divergent lower moments
tends to one of the L\'evy stable class \cite{gnedenko}.

 It is possible to simulate the anomalous behavior of the
 diffusion applying  generalizations on the ordinary diffusion equation.
It can be performed by introducing an appropriate  time-dependence
 \cite{lillo,malatemporal} or spatial dependence
\cite{procaccia1, procaccia2,malanosso,isabel}
 in the equation's coefficients. Also we can   apply fractional derivatives
\cite{ACompte,metzler,bologna,fractional}. However, the
introduction of nonlinearities reveals a large set of
possibilities to describe anomalous diffusive processes.
 An interesting  characteristic of the nonlinear Fokker-Planck equation is that
 its stationary solutions, and some particular time-dependent solutions,
    are such that maximize the Tsallis entropy, a
   nonextensive  entropic form proposed in the last years
by Tsallis \cite{tsallis88,curado}. This effort becomes necessary
when the Boltzmann-Gibbs statistics fails, for instance in the
presence of long range interactions or memory effects and fractal
phase space structure (see Ref. \cite{tsallisgeral} for recent
review). The nonextensive  statistical mechanics  has shown a very
fruitful scenario to study anomalous diffusion. In this way,
 several works
dealing anomalous diffusion were developed in such context. The
connection  between this formalism and the nonlinear Fokker-Planck
equation was first pointed by  Plastino and Plastino
\cite{plastino95}. The results were enlarged by Tsallis and Bukman
\cite{buckman}, including a linear drift term. A phenomenological
microscopic dynamics of the nonlinear Fokker-Planch equation is
presented in \cite{lisa98}, as well a nonlinear Fokker-Planch
equation with state-dependent diffusion\cite{lisa99}. A Tsallis
MaxEnt solutions of the nonlinear Fokker-Planch equation applied to
the study of correlated anomalous diffusion \cite{compte1,compte2},
nonlinear fractional derivative Fokker-Planch like equation
\cite{bologna,fractional}, aging in in nonlinear diffusion
\cite{stariolo}, anomalous diffusion with absorption \cite{drazer},
anomalous diffusion in a fractal dimension \cite{malanosso,isabel},
are some examples among other references.
 In this direction, the generalized thermostatistic, based on the nonextensive
Tsallis
  entropy, becomes a natural scenario
  as the correlated anomalous diffusion  as the anomalous diffusion
  like L\'evy(variance  not finite) \cite{Alemany,zanette95,levy,dprato}.
In this context, the index of nonlinearity joined to the nonlinear
Fokker-Planck equation, in the correlated anomalous diffusion
case,
  can be connected to the entropic index
$q$ of the Tsallis entropy and with the respective generalized
distributions.  This relation is just $q=2-\nu$, where $\nu$ is the
nonlinear index,  and it is easy to verify that when  $q=2$ the
diffusive term becomes trivial. We will show that this difficulty
can be overcame by  a logarithmic diffusion equation, or, in another
words, this equation represents an alternative for the limit case
where the exponent that characterizes
 the nonlinearity of the diffusive term goes to zero.
  In this way, the logarithmic diffusion equation is just inspired in  nonlinear
  diffusion equations like Fokker-Planck  \cite{isabel,plastino95,buckman}.
 We will show that the non-stationary solution for the logarithmic equation
 is a lorentzian,  and this implies that the second moment is not finite. It indicates
 that this diffusion equation is related to superdiffusive
 processes, of the L\'evy kind.
On the other hand, in the nonextensive  statistical mechanics
scenario is possible to unify the  correlated anomalous diffusion
equation, or porous media equation, and the logarithmic diffusion
equation. Thus we enlarge the description of diffusive processes by
a unified equation that congregates the correlated anomalous
diffusion  and a like L\'evy one.

This work is structured as follow.  In the section II we will
present the logarithmic diffusion equation and its exact solution.
 In the section III
this equation is solved including a linear drift term.  In the
following section it is solved when a absorption term is present.
We will obtain in the section V the stationary solution, and in
the section VI we will perform the unification of the porous media
and logarithmic diffusion equations. It is included the connection
with the diffusion on fractals. Finally, in the last section will
be presented the conclusion and final remarks.

\section{Logarithmic Diffusion Equation  }
 To motivate our discussion, we will first  consider
the porous media equation,
 \begin{equation} \label{bd22}
 \frac{\partial \rho}{\partial t}=D \nabla^2 \rho^\nu .
\end{equation}
It has been employed in the analysis of percolation of gases through
porous media ($\nu \geq 2$) \cite{gases}, thin saturated regions in
porous media ($\nu=2$) \cite{thin}, a standard solid-on-solid model
for surface growth ($\nu=3$), thin liquid films spreading under
gravity ($\nu=4$)  \cite{buck}, among others \cite{kath}. A solution
of Eq. (\ref{bd22})  is the generalized $q$-gaussian \cite{buckman}
,
\begin{equation}
\label{bd66} \rho (r,t) = \frac{1}{ Z(t)}{\left[1- (1-q)\,\beta
(t)\, r^2\right]^{\frac{1}{1-q}}}.
\end{equation}
By direct substitution of  (\ref{bd66}) into (\ref{bd22}) it is
easy to verify that
\begin{eqnarray}
 \frac{d \beta}{d t}&=& - 4D\nu\beta^2Z^{q-1}\\
\frac{1}{Z}\frac{dZ}{d t} &=& 2D\beta \nu Z^{q-1}
\end{eqnarray}
and we obtain $\beta(t)\propto t^{-2/(3-q)}$ and $Z(t)\propto
t^{1/(3-q)}$. Of course, when $q\rightarrow 1$ we recover the
results for the usual diffusion. This solution is obtained taking
account the relation  $q=2-\nu$. And what happens when  $q=2$? It is
obvious that in this case the right side of the equation
(\ref{bd22}) vanishes all the time and (\ref{bd66}) is not the
solution now. However, Eq. (\ref{bd66}), for $q=2$, recovers a very
important  function: the lorentzian. A question arises: how would be
the shape of the nonlinear diffusion equation whose solution was a
lorentzian?

It is known that  $\ln x$ decrease more slowly than any  power
 $x^r$, when $r$ goes to zero positively, so we are influenced
 to substitute  $\rho^{\nu}$ by $\ln \rho$  in
Eq. (\ref{bd22}) when $\nu \rightarrow 0$. In this perspective, we
introduce the logarithmic diffusion equation
\begin{equation}
\frac{\partial \rho}{\partial t}=D \frac{\partial^2\ln
\rho}{\partial x^2} .\label{log1}
\end{equation}
This equation emerges in plasma physics \cite{berrymanplasma} and,
in particular, has been predicted for cross-field convective
diffusion of plasma including mirror effects \cite {kamimura}. The
same equation describes the expansion of a thermalized electron
cloud \cite{hirose} and also arises in studies of the central
limit approximation to Carleman's model of the Boltzmann equation
\cite{kurtz,mckean}.

It is verified directly that Eq. (\ref{log1})  presents the
lorentzian
 solution
\begin{equation}
\rho(x,t)=\frac{1}{Z(t)}\frac{1}{[1+\beta(t)x^2]}. \label{log2}
\end{equation}
The substitution  of (\ref{log2}) in (\ref{log1}) shows  that
 $\beta(t)$ and $Z(t)$ obey  the equations
\begin{eqnarray}
\frac{1}{Z^2}\frac{d Z}{d t}&=&2D\beta\label{log3}\\
\frac{1}{Z^2}\frac{d Z}{d t}+ \frac{1}{\beta Z}\frac{d \beta}{d
t}&=&-2D\beta.\label{log4}
\end{eqnarray}
They can be decoupled taking account the relation
$Z\beta^{1/2}=Z_0\beta_0^{1/2}$ which is valid all the time. The
solutions are
\begin{equation}
\beta(t)=\beta_0(1+2D\beta_0Z_0 t)^{-2}\label{log5}
\end{equation}
and
\begin{equation}
 Z(t)=Z_0(1+2D\beta_0Z_0 t).\label{log6}
\end{equation}
It is interesting to point that for the lorentzian the
variance is not finite.  This fact,  characteristic of the L\'evy
distributions, indicates  that the logarithmic equation is
associated to superdiffusive regimes.  In fact, a dimensional
analysis of the logarithmic diffusion equation (\ref{log1}) and
the L\'evy diffusion equation $\partial \rho/ \partial t = D
\partial^\mu \rho /\partial |x|^{\mu}$, with $\mu=1$, leads to the common
ballistic behavior in the sense that $x$ scales as $t$.
 By other side,
 Eq. (\ref{log1}) is a particular $\mu=1$ case of
\begin{equation}
\frac{\partial \rho}{\partial t}=\frac {\partial}{\partial x}\rho
^{-\mu}\frac{\partial \rho}{\partial x} \hskip 2 cm 0<\mu <2.
\end{equation}
When the coefficient diffusion $D(\rho)$ (or the thermal
coefficient of conductivity) can be approximated as $ \rho
^{-\mu}$, its divergence for small $\rho$ causes a much faster
spread of mass (or heat) than in the linear case ($\mu=0$),
justifying  the terminology superfast diffusion to these
processes \cite{rosenau}.

For the $N$-dimensional case,  Eq. (\ref{log1}) assumes the shape
\begin{equation}
\frac{\partial \rho}{\partial t}=D \nabla ^2 \ln \rho  ,
\label{logndim}
\end{equation}
with $\nabla ^2=\sum_{i=1}^N\partial ^2/\partial x_i^2$ and whose
 lorentzian solution,  $\rho(|{\bf x}|,t)$, presents
\begin{equation}
\beta(t)=\beta_0[1-2(N-2)D\beta_0Z_0~ t]^{2/(N-2)}
\label{coisinha1}
\end{equation}
and
\begin{equation}
Z(t)=Z_0[1-2(N-2)D\beta_0Z_0~ t]^{N/(2-N)}. \label{coisinha2}
\end{equation}

\section{Presence of external forces}

Now we analyze   Eq. (\ref{log1}) describing a process that include
a linear drift term  (external force)  $F(x)$,
\begin{equation}
\frac{\partial \rho}{\partial t}=D \frac{\partial^2}{\partial
x^2}\ln \rho -\frac{\partial}{\partial x}(F\rho).\label{log11}
\end{equation}
For the case $F(x)=k_1-k_2x$, the solution of Eq. (\ref{log11}) is
the drifted lorentzian
\begin{equation}
\rho(x,t)=\frac{1}{Z(t)}\frac{1}{[1+\beta(t)(x-x_0(t))^2]}.
\label{log22}
\end{equation}
 In fact,
  $\beta(t)$,  $Z(t)$, and  $x_0(t)$ obey the equations:
\begin{eqnarray}
\frac{d x_0}{dt}&=&k_1-k_2x_0\label{logforca1}\\
\frac{1}{\beta}\frac{d \beta}{d t}&=&-4D\beta Z
+2k_2\label{logforca2}\\
\frac{1}{Z}\frac{d Z}{d t}&=&2D\beta Z-k_2.\label{logforca3}
\end{eqnarray}
  Equation (\ref{logforca1}) does not depend of the index related to the
  non-linearity, therefore it arises  not only in the usual diffusion  equation but
also
   in the porous media ones $(\nu\neq1)$. Thus, the solution of (\ref{logforca1})
   is
\begin{equation}\label{gravita}
x_0 (t) =\left[ x_0 (0)+ \frac{k_1}{k_2} \left(e^{k_2 t} -1\right)
\right] e^{-k_2 t}.
\end{equation}
In this turn, the solutions of (\ref{logforca2}) and
(\ref{logforca3}) have the form
\begin{equation}
\beta(t)=\beta_0\left[1-g(t)\right]^{-2} \label{logsolution1}
\end{equation}
and
\begin{equation}
Z(t)=Z_0\left[1-g(t)\right], \label{logsolution2}
\end{equation}
where
\begin{equation}
g(t)=\left[\frac{(k_2-2D\beta_0
Z_0)}{k_2}\left(e^{k_2t}-1\right)\right]e^{-k_2t}
\label{logsolution3}
\end{equation}
with $g(0)=0$.
\section{Logarithmic equation with source term}
We can consider  the presence of a time-dependent source term. In
this case,  Eq. (\ref{log1}) is written as
\begin{equation}
\frac{\partial \rho}{\partial t}=D \frac{\partial^2}{\partial
x^2}\ln \rho -\alpha(t)\rho.
\label{fonte1}
\end{equation}
The source term in this equation can be removed by an appropriate
change in the solution
\begin{equation}\label{A1}
\rho ({\bf r},t)= e^{\left( -\int_0^t \alpha(t'){d}t'\right)}
\hat{\rho}({\bf r},t)
\end{equation}
 and  we rewrite the time variable. Thus, the solution of (\ref{fonte1})
 is
\begin{equation}
\nonumber
\rho(x,t)=\frac{1}{Z(\tau(t))[1+\beta(\tau(t))x^2]}\exp\left(-\int_0^t\alpha(t')dt'\right),
%\label{fonte2}
\end{equation}
with $\beta(\tau(t))$ and $Z(\tau(t))$ being of the form
(\ref{log5}) and (\ref{log6}), where $D=1$ and $t$ is replaced by
$\tau(t)=\int_0^t\tilde{D}(t')dt'$, with
$\tilde{D}(t)=D\exp\left(\int_0^t\alpha(t')dt'\right)$. We can to
extend this  solution for the  $N$-dimensional case   by employing
(\ref{coisinha1}) and (\ref{coisinha2}).

%XXXXXXXXXXXXXXXXXXXXXXXXXXXXXXXXXXXXXXXXXXXXXXXXXX
\section{Stationary case}

We define  now the probability density current
\begin{equation}
{J}=F\rho-D\frac{\partial \ln \rho}{\partial x} \label{estalog1}
\end{equation}
so that Eq. (\ref{log11}) represents a continuity equation:
\begin{equation}
\frac{\partial \rho}{\partial t}+\frac{\partial {J}}{\partial
x}=0. \label{estalog2}
\end{equation}

In the stationary case, $d {J}/d x=0$,  and this implies
${J}=cte=0$, since we are applying the condition  that the current
vanishes at infinity.  This fact permits us to write, considering
 $F=-d V/d x$,
\begin{equation}
\frac{d \ln \rho}{d x}=-\frac{1}{D} \frac{d V}{d
x}\rho\label{estalog3}
\end{equation}
whose solution is
\begin{equation}
\rho=\frac{1}{1+\beta V}\;, \label{estalog4}
\end{equation}
where it was assumed $V(0)=0$, $\beta=\rho_0/D$ and
$\rho(0)=\rho_0=1$. Note that the structure  (\ref{estalog4}) is
preserved in the $N$-dimensional case.

The solution (\ref{estalog4}) is the  $q=2$ case of the
generalized exponential
\begin{equation}
\rho=[1-(1-q)\beta V]^{1/(1-q)} \label{estalog5}\;,
\end{equation}
which, in this turn, recovers the Boltzmann distribution in the
limit $q\rightarrow 1$. Furthermore, we remind that the Boltzmann
distribution is the stationary solution of the usual diffusion
equation, so (\ref{estalog4})
 is the analogous for the logarithmic equation.
\section{Unification of porous media and logarithmic equations }

The unification of equations  (\ref{bd22}) and (\ref{logndim}) will
be accomplished by using the generalized logarithmic function, the
$q$-logarithm, defined as
\begin{equation}
\ln_q x=\frac{x^{1-q}-1}{1-q}. \label{unilog1}
\end{equation}
In this way, the unified equation proposed in this work presents
the following structure
\begin{equation}
\frac{\partial \rho}{\partial t}=\bar{D} \nabla^{2}\ln_{q-1} \rho,
\label{unilog2}
\end{equation}
and it is clear that, when $q\rightarrow 2$, the function  $\ln_1$
recovers the logarithmic one. When $q=2-\nu \neq 2$, we have
$\ln_{q-1}=(\rho^{\nu}-1)/\nu$ and Eq. ({\ref{unilog2}) reobtains
the porous media equation (\ref{bd22})  with $D=\bar{D}/\nu$.

We can consider the non linear diffusion equation with radial
symmetry, taking account the spatial-dependence in the diffusion
coefficient, $r^{-\theta}$, and a non integral dimension $d$
\cite{malanosso},
  \begin{equation}
\label{bd4}
 \frac{\partial \rho}{\partial t}= D\tilde{\Delta} \rho^\nu \;,
\end{equation}
with
\begin{equation}
\tilde{\Delta}\equiv r^{-(d-1)}\frac{\partial}{\partial r}
r^{d-1-\theta} \frac{\partial}{\partial r} \label{delta1}.
\end{equation}
With the operator $\tilde{\Delta}$ written like this there are not
restrictions for the possible values for $d$. In this way, $d$ can
be interpreted as a fractal dimension   in an embedding
$N$-dimensional space. Equation (\ref{bd4}), in the $\nu=1$
case, recovers the diffusion equation introduced in Refs.
\cite{procaccia1, procaccia2}. Analogously to (\ref{unilog2}), we enlarge the
applicability domain of Eq. (\ref{bd4}) and we have
\begin{equation}
\frac{\partial \rho}{\partial
t}=\bar{D}\tilde{\Delta}\ln_{q-1}\rho ,\label{logfrac1}
\end{equation}
with $\tilde {\Delta}$ defined by (\ref{delta1}). The $ansatz$
\begin{equation}
\label{bd09} \rho(r,t)=\frac{1}{Z(t)}
\left[1-(1-q)\,\beta(t)\,r^\lambda\right]^{\frac{1}{1-q}}\;,
\end{equation}
with $\lambda=2+\theta$, remains valid and substituted  in  Eq.
(\ref{logfrac1})  conducts to the equations
\begin{eqnarray}
\label{bd110}
 \frac{dZ(t)}{dt} &=& \bar{D} \lambda  d \beta(t)
Z^{q}(t) \nonumber
\\
\frac{d\beta(t)}{dt} &=& -\bar{D} \lambda^2  \beta^2(t)
Z^{q-1}(t).
\end{eqnarray}
Such equations are decoupled and solved and their solutions are
 \begin{equation}\label{A5} \beta(t)=\beta_0[1+A
t]^{-\lambda/[\lambda + d(1-q)]}
\end{equation}
and
\begin{equation}\label{A6}
Z(t)=Z_0[1+ A t]^{d/[\lambda+d(1-q)]},
\end{equation}
with
\begin{equation}
A=\bar{D}\lambda [\lambda+d(1-q)] \beta_0 Z_0^{q-1}, \label{AAA}
\end{equation}
$\beta_0=\beta (0)$ and $Z_0=Z(0)$. Thus we obtain the stretched
lorentzian if $\lambda <2$ and short if $\lambda >2$. In fact, for
$q=2$, the solution is
\begin{equation}
\rho(r,t)=\frac{1}{Z(t)}\left[\frac{1}{1+\beta(t)r^{\lambda}}\right]\;,
\label{logfrac3}
\end{equation}
where
%\begin{eqnarray}
$\beta(t)=\beta_0[1+A t]^{-\lambda/[\lambda - d]}$ and $
 Z(t)=Z_0[1+ At]^{d/[\lambda - d]}$.
%\end{eqnarray}
Observe that when $\theta=0$ and $d=N=1$ we obtain the results
 (\ref{bd66}) for $q\neq 2$, and (\ref{log2}) for $q=2$.

We pointed  that the factor $(2-q)$ implicated  Eqs. (\ref{bd22})
and (\ref{bd4})became  trivial at the value $q=2$. With our strategy
we eliminate this undesirable behavior. From the performed
generalization in this section, we can conjecture a nonlinear
diffusion equation, that presents $\rho^{\nu}$,
 can be extended to $\nu=0$, substituting $\rho^{\nu}$
 by $\ln_{q-1}\rho$, with $q=2-\nu$. Thus, in a general way, we can unify
 all the equations presented in this work by the equation
 \begin{equation}
 \frac{\partial \rho}{\partial t}=\sum
 _{i,j=1}^{N}\frac{\partial}{\partial x_i}\left (D_{ij}\frac{\partial}{\partial x_j}
\ln_{q-1}\rho\right)-\sum _{i=1}^{N}\frac{\partial}{\partial
x_i}(f_i\rho)-\alpha (t)\rho.
\end{equation}

\section{Conclusion}
In this work we enlarged the application domain of the porous media
equation for the $\nu=0$ case and we presented the logarithmic
diffusion equation as an alternative procedure to this limit case.
 Its solution has a lorentzian form and it can  characterize
 superdiffusive processes, of L\'evy kind. A unification of porous media
 and logarithmic diffusion equation is obtained, and in a more general form,
 with the fractal nonlinear diffusion equation. This unified description is performed
 in the nonextensive statistical mechanics  scenario.  This accomplishment represents
 a  progress in the formal description of diffusive processes and  in its
 solutions as well.
   In this
direction, the final proposed equation  interpolates  others ones
with consecrated position in the literature.   It is desirable that
the equations presented in this work, or in particular special cases
of them, show physics situations in which there is competition among
different mechanisms  that generate anomalous diffusion.

\end{document}